\documentstyle[12pt]{article}
\begin{document}

\title{\Large\bf Extended supersymmetry in D=1+1}

\author{R. Amorim\thanks{\noindent e-mail: amorim@if.ufrj.br}~~ and
J. Barcelos-Neto\thanks{\noindent e-mail: barcelos@if.ufrj.br}\\
Instituto de F\'{\i}sica\\ Universidade Federal do Rio de Janeiro\\
RJ 21945-970 - Caixa Postal 68528 - Brasil}
\date{}

\maketitle
\abstract
We extend superspace by introducing an antissymetric tensorial
coordinate. The resulting theory presents a supersymmetry with central
charge. After integrating over the tensorial coordinate, an effective
action describing massive bosons and fermions is explicitely derived
for the spacetime dimension $D=2$. The adopted procedure is simpler
than the Kaluza-Klein one and can suggest an alternative for string
compactifications.

\vfill
\vspace{1cm}
\newpage

With the advent of strings we may conclude that fundamental physical
theories should be formulated in a spacetime dimension higher than
four, say $D=10$ or $D=11$. After some compactication mechanism, we
reach the real word in four dimensions. The main problem of this idea
is that there is a large number of ways that compactifications can be
done, each one corresponding to a different effective theory.
Consequently, it is not clear what the fundamental theory should be.

\medskip 
There is another manner to put forward the interpretation that
strings see, or apparently see, spacetime with more dimensions than
four. Reporting back to bosonic strings, we know the spacetime
they see has dimension 26. However, when one considers the spacetime
also containing fermionic degrees of freedom (supersymmetry) its
critical dimension goes down to 10. Keeping this line of reasoning,
we could consider that superspace has more degrees of freedom than
vector bosons and fermions in order to obtain a lower critical
spacetime dimension. Indeed, in the case of spinning strings, when
one introduces antisymmetric tensor degrees of freedom, we verify
that critical dimension is just four \cite{Barc1}. If we do the same
for superstrings, we find quantization more subtle, but we can show
that supersymmetry is only achieved also for $D=4$  \cite{Barc2}. We
then emphasize that spacetime dimension can be actually four, but the
corresponding superspace contains more degrees of freedom than the
usual bosons and fermions. We stress that this procedure is different
from the Kaluza-Klein one \cite{Kaluza}, where the initial spacetime
dimension is higher than the final one after compactification. As
symmetry groups strongly depend on $D$, Kaluza-Klein theories have to
deal with the symmetry reduction in the compactification procedure.
In our approach, extra coordinates are introduced in an extended
superspace constructed over the usual spacetime (which is not
compactified). Consequently, its symmetry group structure is not
modified anytime.

\medskip
Considering that there is some parallelism between fundamental fields
and coordinates of superspace, the consistency of these ideas leads
us to argue if antisymmetric tensors fields might also exist as
fundamental fields. In fact, we mention that the interest for these
fields dates back almost thirty years ago, starting from the work by
Kalb and Ramond \cite{Kalb}, where antisymmetric tensor fields were
introduced as gauge fields that carry the force among string
interactions. Regardless if this fact is true or not, this interest
has been increasing since then. One of the reasons is due to its
particular structure of constraints, related to the reducibility
condition. As a consequence, the quantization of tensor theories
deserves some additional care when compared with the usual case of
gauge theories of rank one \cite{Kaul}. We also mention that
antisymmetric tensor fields appear as one of the massless solutions
of modern string theories (even when tensor coordinates are not
introduced), in company of photons, gravitons, etc. \cite{Ma}.
Another interesting feature of tensor fields is to provide a possible
mechanism of mass generation for gauge fields. This occurs when they
are coupled in a topological way to vector gauge fields
\cite{Cremmer,Barc3}.

\medskip 
Our purpose in this letter is to study with details an extension of
the supersymmetric theory when antisymmetric tensor coordinates are
included. We show that a consistent extension of the supersymmetric
generator is attained by coupling the independent matrix
$\sigma^{\mu\nu}$ with the translational generator in the tensorial
sector. Although this procedure is formally independent of the
spacetime dimension, we only develop the particular case of $D=2$,
due to its simplicity, by using the scalar superfield formulation.
This leads to a off-shell linear realization of supersymmetry with
central charge~\cite{Rivelles}. 

\medskip
In the usual superfield language, fermionic coordinates are
integrated out to obtain a supersymmetric invariant action based on
the usual spacetime coordinates. We do the same with respect to the
tensor coordinate and discuss the effective action which is obtained.
We left the four dimensional case for a future work, specially the
gauge sector, where algebraic complexities are much bigger
\cite{Barc4}.

\bigskip 
Let us first present an extension of the usual supersymmetry in an
arbitrary spacetime dimension $D$ by taking an enlarged superspace
given by $(x^\mu,\,x^{\mu\nu},\,\theta_\alpha)$, where $x^{\mu\nu}$
is an independent antisymmetric tensor coordinate. The physical
meaning of such tensor coordinates may not be evident, but one can
think that, as in the usual fermionic case, they are introduced in
order to construct extended supersymmetric multiplets. We postulate
that the extended supersymmetry transformations are given by

\begin{eqnarray}
\delta x^{\mu\nu}&=&i\,\bar\xi\sigma^{\mu\nu}\theta
\nonumber\\
\delta x^\mu&=&i\,\bar\xi\gamma^\mu\theta
\nonumber\\
\delta\theta_\alpha&=&\xi_\alpha
\label{1}
\end{eqnarray}

\bigskip\noindent
where $\theta$ is a Majorana spinor (see Appendix for details of the
convention and notation, as well as some identities, we are going to
use in this paper) and $\xi$ is the characteristic parameter of the
supersymmetry transformation (also a Majora spinor).

\medskip
The charge generator for the transformations above is

\begin{equation}
Q_\alpha=\frac{\partial}{\partial\bar\theta_\alpha}
+i\bigl(\gamma^\mu\theta\bigr)_\alpha\frac{\partial}{\partial x^\mu}
+\frac{i}{2}\bigl(\sigma^{\mu\nu}\theta\bigr)_\alpha
\frac{\partial}{\partial x^{\mu\nu}}
\label{2}
\end{equation}

\bigskip\noindent
which leads to the extended supersymmetry algebra

\begin{equation}
\bigl\{Q_\alpha,\bar Q_\beta\bigr\}
=-2i\gamma^\mu_{\alpha\beta}\frac{\partial}{\partial x^\mu}
-i\sigma^{\mu\nu}_{\alpha\beta}\frac{\partial}{\partial x^{\mu\nu}}
\label{3}
\end{equation}

\bigskip
\noindent characterizing an additional translation along the
antissymetric tensor ``directions".

\medskip 
These results are valid in any spacetime dimension, but from now on
we are going to stay in $D=2$ due to its simplicity. In this
particular spacetime dimension, the charge generator and the
supersymmetry algebra turn to be

\begin{eqnarray}
&&Q_\alpha=\frac{\partial}{\partial\bar\theta_\alpha}
+i\bigl(\gamma^\mu\theta\bigr)_\alpha\frac{\partial}{\partial x^\mu}
-\frac{1}{2}\bigl(\gamma_5\theta\bigr)_\alpha
\frac{\partial}{\partial y}
\label{3a}\\
&&\bigl\{Q_\alpha,\bar Q_\beta\bigr\}
=-2i\gamma^\mu_{\alpha\beta}\frac{\partial}{\partial x^\mu}
+\gamma_{5\alpha\beta}\frac{\partial}{\partial y}
\label{3b}
\end{eqnarray}

\bigskip\noindent
where we have denoted by $y$ the independent coordinate $x^{01}$.
The part of $Q_\alpha$ with $\gamma_5$ leads to the central charge.
So, antisymmetric coordinates constitute an alternative procedure of
introducing central charge in linear supersymmetry~\cite{Rivelles}
and gives it a well defined physical meaning.

\medskip 
Looking at the expressions for the charge operator and the
supersymmetry algebra in $D=2$, we could, at first sight, conclude
that superspace in $D=2$ with a tensor coordinate, is equivalent to
the usual superspace in $D=3$. This is nonetheless true because
$y=x^{01}$, in $D=2$, is a Lorentz invariant quantity. Consequently,
it could not be identified with $x^2$ coordinate in $D=3$.

\medskip 
Let us now write a supersymmetric Lagrangian for this extended
superymmetric transformation. The best way to do this is by starting
from the superfield notation. The general form of a scalar and real
superfield in $D=2$ reads

\begin{equation}
\Phi=\phi+\bar\theta\psi+\frac{1}{2}\bar\theta\theta F
\label{4}
\end{equation}

\bigskip\noindent 
where the component fields $\phi$, $\psi$ and $F$ depend on $x^\mu$
and $y$. The quantity $\Phi$ is actually a superfield if it
transforms as

\begin{equation}
\delta\Phi=\bigl(\bar\xi Q\bigr)\,\Phi
\label{5}
\end{equation}

\bigskip\noindent
which leads to the following transformations for the component fields
(all of them real)

\begin{eqnarray}
\delta\phi&=&\bar\xi\psi
\nonumber\\
\delta\psi_\alpha&=&\xi_\alpha F
-i(\gamma^\mu\xi)_\alpha\frac{\partial\phi}{\partial x^\mu}
+\frac{1}{2}(\gamma_5\xi)_\alpha\frac{\partial\phi}{\partial y}
\nonumber\\
\delta F&=&\bar\xi\Bigl(-i\gamma^\mu\frac{\partial}{x^\mu}
+\gamma_5\frac{\partial}{y}\Bigr)\psi
\label{6}
\end{eqnarray}

\bigskip
The simplest supersymmetric action with the superfield (\ref{4})
reads

\begin{equation}
S=-\,\frac{1}{2}\int dx^2dyd^2\theta\,\bar D\Phi D\Phi
\label{7}
\end{equation}

\bigskip\noindent
where the derivative operator $D_\alpha$, that anticommute with
$Q_\alpha$, is given by  

\begin{equation}
D_\alpha=\frac{\partial}{\partial\bar\theta_\alpha}
-i\,\bigl(\gamma^\mu\theta\bigr)_\alpha\,
\frac{\partial}{\partial x^\mu}
+\frac{1}{2}\bigl(\gamma_5\theta\bigr)_\alpha\,
\frac{\partial}{\partial y}
\label{8}
\end{equation}

\bigskip\noindent
In the usual case of supersymmetric theories, in superfield language,
we integrate over the Grassmannian coordinates $\bar\theta_\alpha$
and $\theta_\alpha$ to obtain the effective action in component
fields. In the present case, we have also to integrate out the
coordinate $y$. For the integration over the Grassmannian
coordinates, we get

\begin{equation}
S=-\frac{1}{2}\int d^2xdy\,
\Bigl(i\bar\psi\gamma^\mu\frac{\partial\psi}{\partial x^\mu}
-\frac{1}{2}\bar\psi\gamma_5\frac{\partial\psi}{\partial y}
+\frac{\partial\phi}{\partial x^\mu}
\frac{\partial\phi}{\partial x_\mu}
-\frac{1}{4}\frac{\partial\phi}{\partial y}
\frac{\partial\phi}{\partial y}+F^2\Bigr)
\label{9}
\end{equation}

\bigskip\noindent
One can easily verify that indeed Eq. (\ref{9}) is invariant under
transformations (\ref{6}). 

\medskip 
With respect the $y$-coordinate, we consider it is not an unlimited
variable like $x^\mu$, but defines a circle of radius $R$. A
consistency for this hypothesis is related to the fact that $y$ is an
invariant quantity and, consequently, there is no problem related
with Lorentz transformations. At this point we realize one of the
problems of dealing with a spacetime of higher dimensions
\cite{Barc4}. Let us consider the Fourier expansion

\begin{equation}
\phi^A(x,y)=\frac{1}{\sqrt R}\sum_{n=-\infty}^{\infty}
\phi^A_n(x)\exp\Bigl(2in\pi\frac{y}{R}\Bigr)
\label{10}
\end{equation}

\bigskip\noindent
where $\phi^A$ is representing any of the component fields $\psi$,
$\phi$, and $F$. Since all of the component fields are real, we have
that the Fourier modes must satisfy the condition

\begin{equation}
\phi^A_n(x)=\phi^{A\ast}_{-n}(x)
\label{11}
\end{equation}

\bigskip\noindent 
Introducing the Fourier expansion given by (\ref{10}) into expression
(\ref{9}) and performing the integration over $y$, we get an
effective action in usual space-time spanned by the vector coordinate
$x^\mu$ 

\begin{eqnarray}
S&=&-\frac{1}{2}\sum_{n=0}^\infty\int d^2x\,
\Bigl(i\bar\psi_n^\ast\gamma^\mu\partial_\mu\psi_n
-i\frac{n\pi}{R}\bar\psi_n^\ast\gamma_5\psi_n
\nonumber\\
&&\phantom{-\frac{1}{2}\sum_{n=0}^\infty\int d^2x\,\Bigl(}
-\phi_n^\ast\Box\phi_n
-\frac{n^2\pi^2}{R^2}\phi_n^\ast\phi_n
+F_n^\ast F_n\Bigr)
\label{12}
\end{eqnarray}

\bigskip\noindent 
It is interesting to note that for non zero modes both bosonic and
fermionic fields components  are massive. As it is well known, a
massive Dirac fermion component satisfies the massive Klein-Gordon
equation as a consequence of the iteration of Dirac equation. So, if
$(i\slash\!\!\!\partial+m)\psi=0$, then $(i\slash\!\!\!\partial
+m)(i\slash\!\!\!\partial-m)\psi=-(\Box+m^2)\psi=0$. In our case, the
mode of order $n$ with mass $m=n\pi/R$ satisfies the equation
$(i{\slash\!\!\!\partial}-im\gamma_5)\psi=0$, which also implies
under iteration that
$(i\slash\!\!\!\partial-im\gamma_5)^2\psi=-(\Box+m^2)\psi=0$. This is
an expected result, since we verify that the equations of motion for
the fermionic massive modes are just the usual Dirac equation with
the Dirac matrices written with the aid of a similarity
transformation given by the unitary matrix
$U=\exp(i\frac{\pi}{4}\gamma_5)$.

\medskip
The action above is invariant under

\begin{eqnarray}
\delta\phi_n&=&\bar\xi\psi_n
\nonumber\\
\delta\psi_n&=&F_n\xi
-i\partial_\mu\phi_n\gamma^\mu\xi
+\frac{in\pi}{R}\phi_n\gamma_5\xi
\nonumber\\
\delta F_n&=&-i\bar\xi\gamma^\mu\partial_\mu\psi_n
+\frac{in\pi}{R}\bar\xi\gamma_5\psi_n
\label{13}
\end{eqnarray}

\bigskip\noindent 
Observe that the zero modes satisfy the usual supersymmetry, but
extended supersymmetries depending on the mass parameters $n\pi/R$
arise when the other modes are taken in consideration. Although its
unusual origin (superspace with tensor degrees of freedom), the
effective theory describes an infinite set of bosons and fermions
with the usual behavior in $D=2$.  

\vspace{1cm}
\noindent {\bf Acknowledgment:} This work is supported in part by
Conselho Nacional de Desenvolvimento Cient\'{\i}fico e Tecnol\'ogico
- CNPq, Financiadora de Estudos e Projetos - FINEP, and
Funda\c{c}\~ao Universit\'aria Jos\'e Bonif\'acio - FUJB (Brazilian
Research Agencies). 

\vspace{1cm}
\appendix
\renewcommand{\theequation}{A.\arabic{equation}}
\setcounter{equation}{0}
\section*{Appendix}

\bigskip
In this Appendix, we present the notation, convention and the main
identities we use in the paper. The gamma matrices satisfy the usual
relations 

\begin{eqnarray}
&&\{\gamma^\mu,\gamma^\nu\}=2\,\eta^{\mu\nu}
\nonumber\\
&&\gamma^\mu=\gamma^0\gamma{\mu\dagger}\gamma^0
\label{A.1}
\end{eqnarray}

\bigskip\noindent 
We adopt the metric convention $\eta^{\mu\nu}=diag.\,(1,-1)$ and the
following representation for the gamma matrices

\begin{equation}
\gamma^0=\sigma_2=\left(\begin{array}{cc}
0&-i\\
i&0\\
\end{array}\right)
\hspace{1cm}
\gamma^1=i\sigma_1=\left(\begin{array}{cc}
0&i\\
i&0\\
\end{array}\right)
\label{A.2}
\end{equation}

\bigskip\noindent
They also satisfy the relation

\begin{equation}
\gamma^\mu\gamma^\nu=\eta^{\mu\nu}\,1
+\epsilon^{\mu\nu}\,\gamma_5
\label{A.3}
\end{equation}

\bigskip\noindent
with the antisymmetric quantity $\epsilon^{\mu\nu}$ defined by
$\epsilon^{01}=1$ and 

\begin{equation}
\gamma_5=\gamma^0\gamma^1=\left(\begin{array}{cc}
1&0\\
0&-1\\
\end{array}\right)
\label{A.4}
\end{equation}

\bigskip\noindent
Consequently, 

\begin{eqnarray}
\sigma^{\mu\nu}&=&\frac{i}{4}\bigl[\gamma^\mu,\gamma^\nu\bigr]
\nonumber\\
&=&\frac{i}{2}\,\epsilon^{\mu\nu}\gamma_5
\label{A.5}
\end{eqnarray}

\bigskip\noindent 
Before Fourier expansion, all the spinors that appear in the text are
Majorana and in the Majorana representation, what means that they are
real. Let us list below some of useful identities that the Majorana
spinors satisfy

\begin{eqnarray}
\bar\psi\chi&=&\bar\chi\psi
\nonumber\\
\bar\psi\gamma^\mu\chi&=&-\bar\chi\gamma^\mu\psi
\nonumber\\
\bar\psi\gamma_5\chi&=&-\bar\chi\gamma_5\psi
\nonumber\\
\bar\psi\gamma^\mu\gamma^\nu\chi
&=&-\bar\chi\gamma^\nu\gamma^\mu\psi
\nonumber\\
\psi_\alpha\psi_\beta
&=&-\frac{1}{2}(\gamma^0)_{\alpha\beta}\bar\psi\psi
\label{A.6}
\end{eqnarray}

\bigskip\noindent 
We also list some other relations involving derivatives and
integrations over Berezin coordinates:

\begin{eqnarray}
&&\frac{\partial}{\partial\theta_\alpha}\,\bar\theta_\beta
=\gamma^0_{\alpha\beta}
=\frac{\partial}{\partial\bar\theta_\alpha}\,\theta_\beta
\nonumber\\
&&\frac{\partial}{\partial\theta_\alpha}\,(\bar\theta\theta)
=-2\bar\theta_\alpha
\nonumber\\
&&\frac{\partial}{\partial\bar\theta_\alpha}\,(\bar\theta\theta)
=2\theta_\alpha
\nonumber\\
&&d^2\theta\equiv\frac{1}{4}\,d\bar\theta d\theta
\nonumber\\
&&\int d^2\theta\,\bar\theta\theta=1
\label{A.7}
\end{eqnarray}


\vfill
\newpage 
 
\begin{thebibliography}{30} 
\bibitem{Barc1} R. Amorim and J. Barcelos-Neto, Z. Phys. C58
(1993)513. 
\bibitem{Barc2} R. Amorim and J. Barcelos-Neto, Z. Phys. C64
(1994)345. 
\bibitem{Kaluza} T. Kaluza, Akad. Wiss. Phys. Math. K1 (1921) 966; O.
Klein, Z. Phys. 37 (1926) 895. See also L. Castellani, R. D'Auria and
P. Fr\'e, {\it Supergravity and Superstrings -- A Geometric
Perspective} (World Scientific, 1991), and references therein.
\bibitem{Kalb} M. Kalb and P. Ramond, Phys. Rev. D9 (1974) 2273.
\bibitem{Kaul} See, for example, R.K. Kaul, Phys. Rev. D18 (1978)
1127. C.R. Hagen, Phys. Rev. D19 (1979) 2367; A. Lahiri, Mod.
Phys. Lett. A8 (1993) 2403; J. Barcelos-Neto and M.B.D. Silva, Int.
J. Mod. Phys. A10 (1995) 3759; Mod. Phys. Lett. A11 (1996) 515. R.
Banerjee and E.C. Marino, Nucl. Phys. B507 (1997) 501; R. Banerjee
and J. Barcelos-Neto,  Ann. Phys. 265 (1998) 134. See also M.
Henneaux and C.  Teitelboim, {\it Quantization of gauge systems}
(Princeton University Press, New Jersey, 1992) and J. Gomis, J.
Par\'is and S. Samuel, Phys. Rep. 259 (1995) 1, and references
therein.
\bibitem{Ma} J. Maharana and J.H. Schwarz, Nucl. Phys. B390 (1993) 3.
\bibitem{Cremmer} E. Cremmer and J. Scherk, Nucl Phys. B72 (1974)
117.  See also, T.J. Allen, M.J. Bowick and A. Lahiri, Mod. Phys.
Lett. A6 (1991) 559. 
\bibitem{Barc3} R. Amorim and J. Barcelos-Neto, Mod. Phys. Lett.
A10 (1995) 917; J. Barcelos-Neto and S. Rabello, Z. Phys. C74 (1997)
7.
\bibitem{Rivelles} See for example P. Barbosa and V.O. Rivelles,
Nucl. Phys. B271 (1985) 80, where this kind of realization is
achieved in a non-linear way.
\bibitem{Barc4} R. Amorim and J. Barcelos-Neto, work in progress.
\end{thebibliography}
\end{document}